\documentclass[12pt,preprint]{aastex}

\usepackage{graphicx}
\usepackage{amssymb}
\usepackage{natbib}
\usepackage{amsmath}
\usepackage{url}

\slugcomment{}

\shorttitle{Case study of four homologous large-scale coronal waves observed on April 28-29 2010}
\shortauthors{Kienreich et al.}

\begin{document}

\title{Case study of four homologous large-scale coronal waves observed on April 28-29 2010}

\author{I.W. Kienreich}
\affil{Institute of Physics, University of Graz,
    Universit\"atsplatz 5, A-8010 Graz, Austria}
    \email{ines.kienreich@uni-graz.at}

\author{A.M. Veronig}
\affil{Institute of Physics, University of Graz,
    Universit\"atsplatz 5, A-8010 Graz, Austria}

\author{N. Muhr}
\affil{Institute of Physics, University of Graz,
    Universit\"atsplatz 5, A-8010 Graz, Austria}

\author{M. Temmer}
\affil{Institute of Physics, University of Graz,
    Universit\"atsplatz 5, A-8010 Graz, Austria;
    Space Research Institute, Austrian Academy of Sciences, Schmiedlstra{\ss}e 6, A-8042 Graz, Austria
    }

\author{B. Vr\v{s}nak}
\affil{Hvar Observatory, Faculty of Geodesy, Ka\v{c}i\'ceva 26, 1000 Zagreb, Croatia}

\author{N. Nitta}
\affil{LMSAL, 3251 Hanover Street, Palo Alto, CA 94304 USA}

\begin{abstract}
On April 28--29 2010, the Solar Terrestrial Relations Observatory (STEREO)-B/Extreme Ultraviolet Imager (EUVI) observed four homologous large-scale coronal waves, so-called `EIT-waves', within eight hours. All waves emerged from the same source active region, were accompanied by weak flares and faint CMEs, and propagated into the same direction at constant velocities in the range of $\sim220-340\,\mathrm{km\,s}^{-1}$. The last of these four coronal wave events was the strongest and fastest, with a velocity of $337 \pm 31\,\mathrm{km\,s}^{-1}$ and a peak perturbation amplitude of $\sim1.24$, corresponding to a magnetosonic Mach number of $M_{\mathrm{ms}} \sim1.09$. The magnetosonic Mach numbers and velocities of the four waves are distinctly correlated, suggestive of the nonlinear fast-mode magnetosonic wave nature of the events. We also found a correlation between the magnetic energy build-up times and the velocity and magnetosonic Mach number.
\end{abstract}



\section{Introduction}

Large-scale propagating disturbances in the solar corona have been subject of extensive studies for more than twelve years. For the first time, these events were observed by the Extreme-ultraviolet Imaging Telescope \citep[EIT;][]{delaboudiniere95} onboard the Solar and Heliospheric Observatory \citep[SoHO; ][]{moses97,thompson98}, and are thus commonly called `EIT-waves'. They appear as diffuse coronal brightenings, forming circular wave fronts, traveling hundreds of Mm without hindrance under quiet-Sun conditions.

At present there are two competing groups of models based on completely different ideas of the physical nature of EIT, or more generally, EUV waves. The first theory describes them as nonlinear fast-mode magnetosonic waves following the original interpretation of large-scale disturbances by Uchida \citep[e.g.][]{thompson98,wills99,wang00,klassen00,wu01,vrsnak02,warmuth04a,veronig06}. Several characteristics of coronal waves can be explained by this theory: (1) the propagation perpendicular to magnetic field lines \citep[e.g.][]{thompson98}, (2) the pulse broadening and amplitude drop off \citep{wills06,warmuth10,veronig10},  (3) the reflection and refraction at regions of high Alfv\'en velocity \citep{thompson99,veronig08,long08,gopalswamy09}. The second group of models, however, consider these coronal bright fronts as no waves at all, but as the signature of a large-scale magnetic reconfiguration of field lines during a CME lift-off \citep[e.g.][]{delanee99,chen02,attrill07}. In these pseudo-waves models, coronal waves could exclusively occur in conjunction with CMEs. This assumption is supported by statistical studies, showing a close relation between waves and CMEs \citep[e.g.][]{biesecker02}. These models can also explain stationary brightenings \citep{delanee99,cohen09} and the occasionally observed rotation of coronal bright fronts \citep{podladchikova05}. For recent reviews we refer to \citet{vrsnak08}, \citet{wills09} and \citet{warmuth10}.

The limitations of EIT observations, especially the low imaging cadence of $\sim$12--15 minutes, were a major contributor to the difficulties determining the physical nature of coronal waves. The Extreme Ultraviolet Imager \citep[EUVI;][]{wuelser04} instruments, which are
part of the Sun Earth Connection Coronal and Heliospheric Investigation \citep[SECCHI;][]{howard08} suite onboard the twin Solar-TErrestrial Relations Observatory \citep[STEREO;][]{kaiser08} spacecraft overcome these limitations. EUVI observes the entire solar disk as well as the corona up to 1.7 R$_{\odot}$ in four spectral channels (He II 304\,\AA: T $\sim$ 0.07 MK; Fe IX 171\,\AA: T $\sim$ 1 MK; Fe XII 195\,\AA: T $\sim$ 1.5 MK; Fe XV 284\,\AA: T $\sim$ 2.25 MK). The high imaging cadence down to $\sim75$ seconds, the large FoV as well as the ability of simultaneous observations from two vantage points, provide us with new insights \textbf{into} the three-dimensional kinematics and dynamics of EUV waves \citep[e.g.][]{kienreich09,patsourakos09,veronig10}.

In this letter we present the first case of homologous EIT waves ever reported in the nearly 14 years of continuous studies of large-scale coronal waves. The four homologous waves, observed by STEREO-B within a period of eight hours, were launched from the same AR, propagated in the same direction, and their wave fronts were similar in both shape and angular extent. These events allow us to perform a detailed study of the physical characteristics of large-scale coronal waves, as these homologous waves are initiated and propagate under similar coronal background conditions. Hence, our analysis of the correlation between the wave pulse parameters is not influenced by the limiting factor of differing coronal plasma conditions, persistent in previous comparative studies of EIT waves.

\section{Data}

 The events under study occurred on 2010 April 28--29, and were observed by STEREO-B (henceforth ST-B), positioned 70.5\degr\ behind Earth on its orbit around the Sun. For our analysis, only the high-cadence EUVI imagery in 195\,\AA\ (cadence of 5 minutes) was suitable, since the 171\,\AA\ and 284\,\AA\ filtergrams were recorded only every 2 hours, and the 304\,\AA\ data revealed no wave signatures at all. In addition, the white-light coronagraph (WLC) images of the COR1-B and COR2-B instruments \citep[][]{howard08} and of the Large Angle and Spectrometric COronagraph \citep[LASCO;][]{brueckner95} C2 instrument onboard SOHO were used to identify signatures of associated CMEs. The EUVI 195\,\AA\ filtergrams were reduced using the SECCHI\_PREP routines available within Solarsoft. Additionally we differentially rotated the images to the same reference time. In order to enhance faint coronal wave signatures, we derived running ratio (RR) images dividing each direct image by a frame taken 10~minutes earlier.

% FIG. 1

\section{Results} \label{results}

\subsection{Wave characteristics}
\begin{figure}[p]
\resizebox{16cm}{!}{\includegraphics[angle=0]{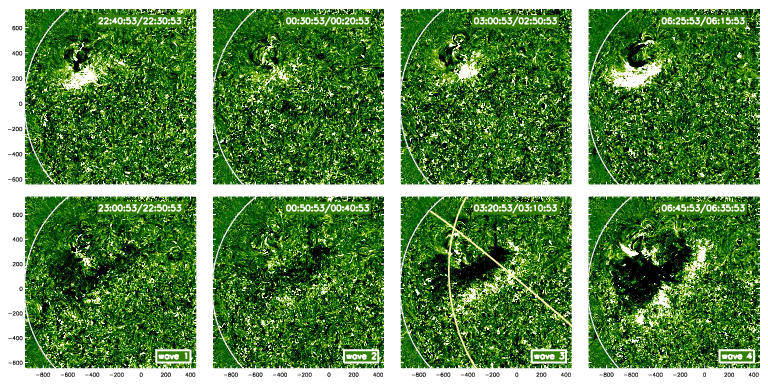}}
\caption{Median-filtered 10 minutes running ratio images recorded in EUVI-B 195\,\AA\  showing the early evolution of four homologous waves on 2010 April 28--29 (axes in arcsec). Each column consists of two images recorded 20~minutes apart of one particular event. Upper panel: wave fronts at the time of the peak perturbation amplitudes [See accompanying movie 1]. Panel 7:
yellow meridians define the sector, for which the wave kinematics was analyzed.} \label{fig1}
\end{figure}

The four EUV waves under study were launched on 2010 April 28--29 within a period of 8 hours (see\ Fig.~\ref{fig1}). Movie 1 shows ST-B 195\,\AA\ direct and running ratio full disk images covering the full observation period. Figure 1 consists of snapshots from the movie illustrating the evolution of all four waves. In each column we present two fronts of each event, with the upper panels showing the wave front with the maximum intensity amplitude, and the lower panels showing the propagating wave front 20 minutes later. All four coronal disturbances are launched from the same AR NOAA 11067, propagate in the same direction and have a similar appearance and angular extent, thus it is appropriate to call them homologous. The time interval between the onsets of the successive waves steadily increased: $\sim$1.75 hr (wave 1--2), $\sim$2.5 hr (wave 2--3) and $\sim$3.5 hr (wave 3--4). In order to obtain better insight into the onset of the coronal waves, we studied a small area [FoV: $600\arcsec\times600\arcsec$] centered on the source AR. Figure~\ref{fig2}(f,g) gives one snapshot of this region during the launch of wave 4, the strongest one of the four events. During the full observation period we recorded persisting dynamical processes in the extended loop system of NOAA 11067 (see movie 2 available as online supplement).

 The resemblance of the events suggests that the background coronal field has not changed noticeably within these eight hours. Moreover, this implies that the corona is disturbed by the EIT waves only for a short period of time and returns to its pre-event condition on a timescale of tens of minutes. In at least two cases, waves 3 and 4, there is evidence that the main perturbation was followed by an additional weak disturbance about 20 minutes later (see movie 1).

%Fig. 2

\subsection{Wave kinematics and perturbation characteristics} \label{wavekin}

We analyzed the kinematics of the waves, each treated as unique event, employing two different methods. Firstly, we visually determined the wave fronts in a series of RR images. Secondly, we examined the perturbation amplitudes of the wave. In both methods we focused on a 60\degr\ sector on the solar surface, where all four waves were distinctly observable (see\ Fig.~\ref{fig1} panel 7; yellow curves).

In the first method the wave fronts were tracked manually, and their center was obtained from a circular fit to the earliest wave fronts carried out in spherical coordinates \citep[see][]{veronig06}. The mean distance of each wave front from the thus determined center is calculated along the solar surface. In the second method we subdivided the solar surface into spherical segments of equal width concentric around the wave center obtained by method one. Plotting the average intensities versus mean distances of all segments gives one intensity profile per RR image \citep[cf.][]{muhr10}. In these perturbation profiles, shown in Fig.~\ref{fig4} (top panels), the wave front presents itself as distinct bump above the background intensity. In each case, the perturbation amplitude reaches its maximum $\sim$5 minutes after the onset of the wave (see Fig.~\ref{fig4}). As example, the evolution of the perturbation amplitude of wave 4 is plotted in Fig.~\ref{fig3} (bottom panel). From the wave perturbation profiles we extracted the foremost position of the wave front, defined as the point at which the gaussian fit to the profile falls below the value of $I/I{_0} = 1.02$ (blue dotted line in top panel of Fig.~\ref{fig4}). We note that the visually tracked distances match well the position of the wave front obtained from the perturbation profiles (Fig.~\ref{fig4}).

In Fig.~\ref{fig3} (top panel) we show the kinematics for wave 4 derived with both methods together with error bars. Linear as well as quadratic least squares fits were applied to the time-distance data, both yielding similar velocities of $\sim340\,\mathrm{km\,s}^{-1}$. The parabolic fit suggests a small deceleration of $-10\,\mathrm{m\,s}^{-2}$. In order to distinguish whether the linear or parabolic fit better represents the data, we derived the 95\% confidence interval for the linear fit. The quadratic fit lies within the error bars and the confidence interval, thus it is reasonable to represent the wave's kinematics by the linear fit with a constant velocity of $337 \pm 31\,\mathrm{km\,s}^{-1}$ over the full propagation distance up to $800\,\mathrm{Mm}$.

We found similar results for the other three wave events under study, which is illustrated in Fig.~\ref{fig4} (bottom panels). The velocities $v_{\mathrm{c}}$ of the four homologous waves are: $257\pm22\,\mathrm{km\,s}^{-1}$ (wave 1), $219\pm18\,\mathrm{km\,s}^{-1}$ (wave 2),  $249\pm18\,\mathrm{km\,s}^{-1}$ (wave 3) and $337\pm31\,\mathrm{km\,s}^{-1}$ (wave 4). All peak perturbation profiles together with a gaussian fit are shown in Fig.~\ref{fig4} (top panel). The perturbation profiles of all four waves are steepening and show an increase in amplitude in the early phase of their evolution until the peak perturbation amplitude $A_{\mathrm{max}}$ is reached. The values $A_{\mathrm{max}}$ of all four waves are: $1.15$ (wave 1), $1.1$ (wave 2), $1.14$ (wave 3) and $1.24$ (wave 4).

The bright fronts of coronal EUV waves are in general caused by a local temperature and density enhancement (plasma compression). Assuming that the change in density is more important than that in temperature, one can derive an estimate of the density jump from the intensity amplitude $A=I/I_0$, $N/N_0 \sim (I/I_0)^{1/2}$. This implies for the peak amplitudes of the four homologous waves a maximum density jump of $X_{\mathrm{c}}= N/N_0$ = $1.07$ (wave 1), $1.05$ (wave 2), $1.07$ (wave 3), and $1.11$ (wave 4).

\subsection{Associated CMEs and flares}

\begin{center}
\begin{figure}[p]
\resizebox{9cm}{!}{\includegraphics{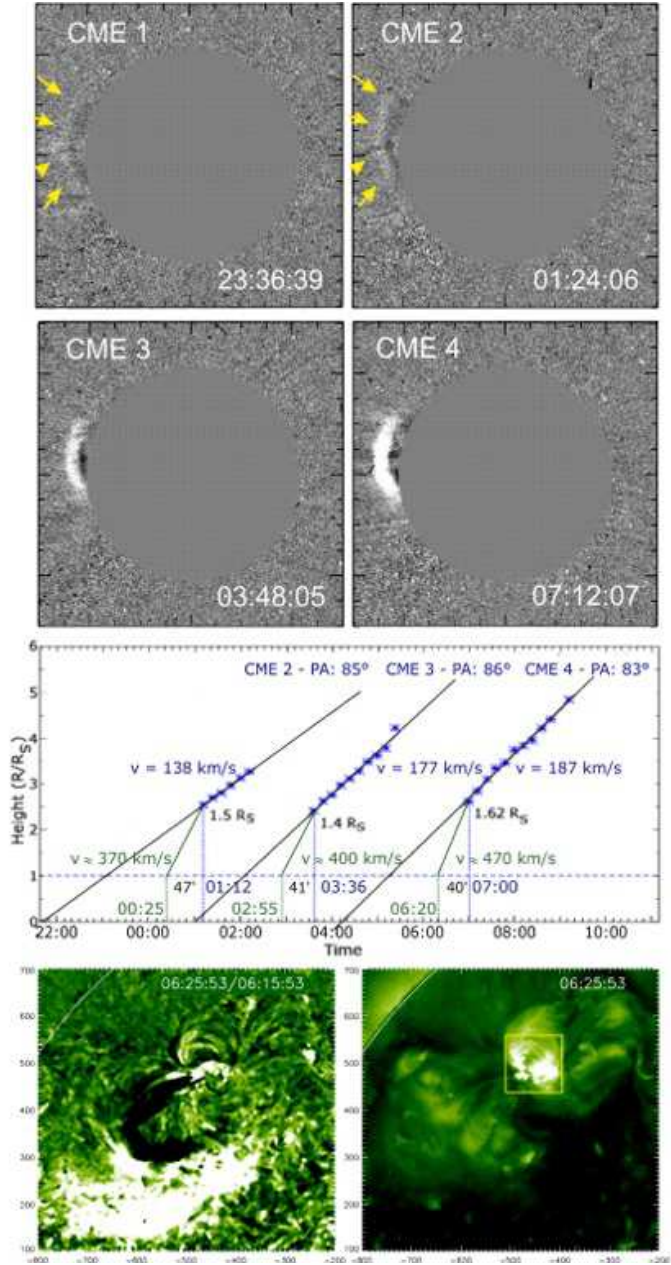}}
\caption{Associated features: (a)-(d) LASCO C2 running difference images of associated weak CMEs at the time of the maximum leading edge intensity [cf. movie 3]. (e) Kinematics of CMEs 2--4 in C2 (blue stars), plus estimated initial CME speed (green lines) by setting the moment of the first observed associated wave front as CME onset time. CME 1 was too faint to be measured. (f) Median-filtered EUVI-B 195\,\AA\ running ratio image at 06:25~UT showing the early phase of wave 4 and the disappearance of an AR loop system (axes in arcsec). (g) Co-temporal direct image revealing an intensity increase of the inner loop system of AR 11067 and the associated flare [cf. movie 2]. yellow square: $110\arcsec\times110\arcsec$ subfield used to analyze flare intensity variations.}\label{fig2}
\end{figure}
\end{center}

\begin{center}
\begin{figure}
\resizebox{12cm}{!}{\includegraphics{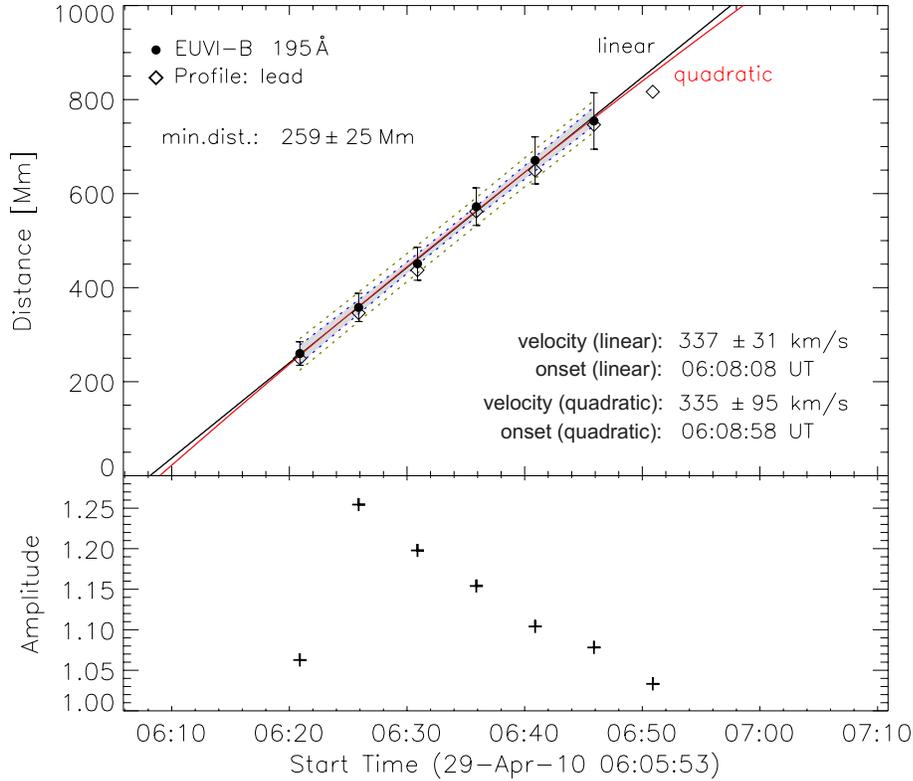}}
\caption{Top: Kinematics of wave 4 determined by two methods plus error bars reflecting the diffusiveness of the wave fronts. Kinematics of the visually determined wave fronts in 195\,\AA\  represented by black circles, of the gaussian fitted perturbation profile by diamonds. A linear (black) and a quadratic (red) least square fit are overlaid. The grey area indicates the 95\% confidence interval of the linear fit. Bottom: evolution of the perturbation amplitude determined from EUVI-B 195\,\AA\ intensity profiles in units of pre-event intensity $I_0$ (compare Fig.~\ref{fig4}, top panels).\label{fig3}}
\end{figure}
\end{center}

\begin{figure}
\resizebox{16cm}{!}{\includegraphics{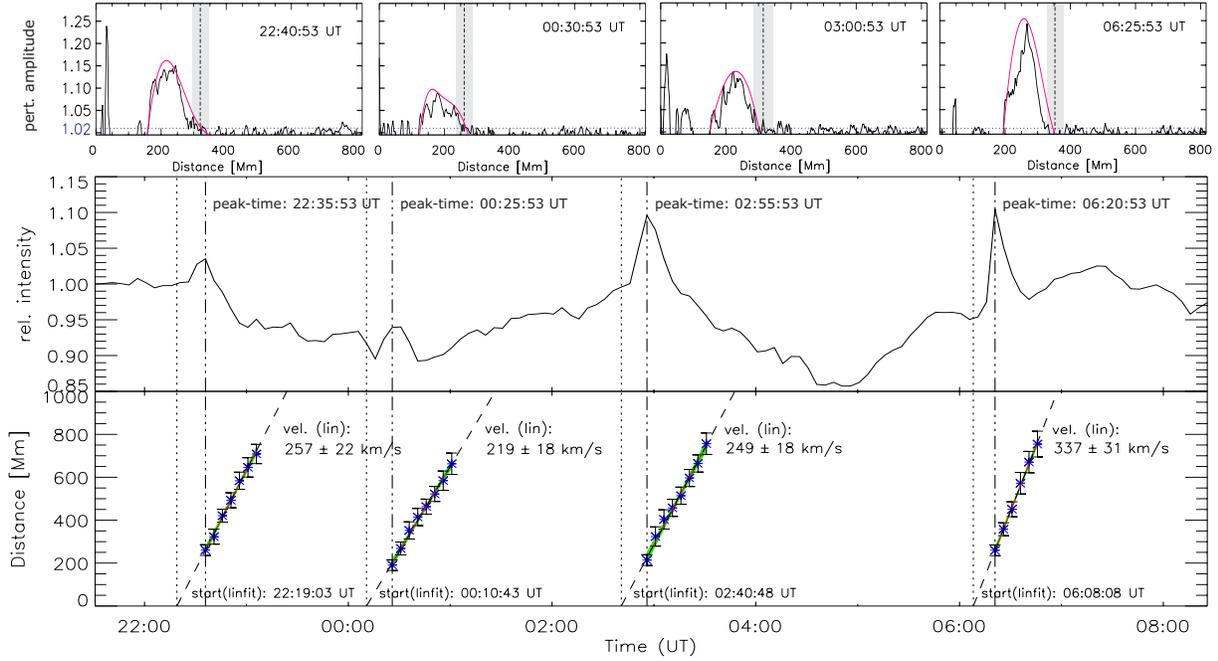}}
\caption{Top: 195\,\AA\ peak intensity enhancements (`perturbation profiles') of the four waves together with Gaussian fits derived from RR images in a 60\degr\ sector (c.f. Fig.~\ref{fig1}; panel 7). Each vertical line indicates the distance of the visually tracked wave front, grey bars represent measured errors. Middle: variations of the flare intensity (in units of pre-event intensity; cf. Fig.~\ref{fig2}(g), yellow square). Bottom: time distance diagram of all waves together with their linear fits. Green areas illustrate the 95\% confidence bands for each linear fit. Note that the first wave fronts coincide with the flare peak intensity of the subfield.\label{fig4}}
\end{figure}

 The analysis of EUVI, COR1 and COR2 images, covering both days of April 28--29, revealed no clear evidence of associated CMEs. However, LASCO C2 recorded four faint CMEs, each entering the C2 FoV about 45 minutes after the first observation of a wave front by EUVI (cf. Fig.~\ref{fig2} (a)-(d) and movie 3). In the LASCO CME catalog just CMEs 2 -- 4 are listed and classified as poor, only visible in C2 at position angles $\sim$85$\degr$. CMEs 3 and 4 reveal additional bright features following the actual CME leading edge with at least the same speed. These features resemble rather small scale ejecta than prominences trailing the leading edge and could be related to the above mentioned weak disturbances following wave 3 and 4 (see movie 1). As the center of the source AR lies $\sim$15$\degr$ behind the limb, as seen from SoHO, we experience low projection effects in the derived CME plane-of-sky velocities, which lie in the range of $~140 - 190\mathrm{\,km\,s}^{-1}$ (see Fig.~\ref{fig2} (e)). Setting the onset site to the solar limb and onset time to the moment, EUVI observed the first front of the associated wave, we estimated for each of CMEs 2--4 an average initial velocity in the range of $~370-470\mathrm{\,km\,s}^{-1}$. We can speculate, that the kinematics shown in Fig.~\ref{fig2} (e) supports the idea of a strong deceleration of the faint, i.e. low-density, CMEs during their early evolution.

In each of the four wave events, an increase in intensity in the central region of the source loop system (see\ Fig.~\ref{fig2}(g)), constituting weak flares, is observed $\sim5$ minutes before the first recorded wave front. Since the wave events occurred behind the solar limb, as viewed from Earth, no GOES X-ray data are available to obtain the flare class. As the determined wave kinematics allows us to back-extrapolate the start of the wave (see\ Fig.~\ref{fig4}; bottom panel), we can compare the timing of the flare commencement and the wave onsets (see also movie 2). Fig.~\ref{fig4} (middle panel) shows the evolution of the total intensity of the central region in the AR. Four distinct intensity peaks are discernible, which coincide with the first observations of the four coronal waves. During these flaring phases we observe the disappearance of several loops in the 195\AA\ RR-images (see\ Fig.~\ref{fig2}(f)), whose northern branch is rooted close to the flaring part of the AR. The coronal waves are, however, launched from the opposite side and the shape of each first wave front exactly maps out the geometry of the southern loop system.

\section{Discussion and Conclusions}

Never before in the nearly 14 years of continuous research of large-scale coronal waves homologous EIT wave events were reported. We are the first to present a study of homologous EIT waves, emerging from the same AR within a short period of time. They travel into the same direction and their fronts have similar shape and angular extent. They propagate into a quiet Sun area, surrounded by ARs to the north and south and a large coronal hole close to the northern polar region (see\ movie 1). As is expected for nonlinear magnetosonic waves, they do not penetrate into these regions of increased Alfv\'en velocity \citep[see also][]{veronig08,gopalswamy09}. In our study we compared for the first time different methods of deriving the wave kinematics, the (rather subjective) visual method and the (more objective) profile method. Both methods yield consistent results, i.e. the waves propagate at constant velocities $\sim220\mathrm{\,km\,s}^{-1}$ for the weakest wave up to $\sim340\mathrm{\,km\,s}^{-1}$ for the strongest event. Furthermore, we calculated the perturbation profiles to study the physical characteristics and evolution of the disturbances. The strong initial steepening of the perturbation amplitudes confirms that these features are indeed shocks, albeit only weak shocks, since they peak at low intensity values $A < 1.25.$

Assuming these coronal waves to constitute large-scale fast magnetosonic waves, the measured velocities lie well within the velocity range of $210 - 350\mathrm{\,km\,s}^{-1}$ for fast magnetosonic waves for quiet Sun conditions. In the MHD approach the quantities defining a shock wave in the solar corona are: $M_{\mathrm{ms}}$ \dots shock magnetosonic Mach number, $X_{\mathrm{c}}$ \dots density jump at the shock front, $\vartheta$ \dots angle between shock front and magnetic field, $\beta_{\mathrm{c}}$ \dots plasma-beta. They are related by the Rankine-Hugoniot (RH) conditions for an oblique shock \citep[c.f.][]{priest82}. Considering a perpendicular shock the RH relation reduces to
\begin{equation*}
    M_{\mathrm{ms}} = \sqrt{\frac{X_{\mathrm{c}}(X_{\mathrm{c}}+5+5\beta_{\mathrm{c}})}{(4-X_{\mathrm{c}})(2+\gamma\beta_{\mathrm{c}})}},
\end{equation*}
with a polytropic index $\gamma$ of $5/3$. Studies of \citet{vrsnak02} indicate that $\beta\approx0.1$ in the quiet Sun's low corona. Using the previously calculated density jumps $X_c$ (cf. Sect. \ref{wavekin}, $X_c \propto\sqrt{A}$), we derive for the peak magnetosonic Mach numbers $M_{\mathrm{ms}}$= $1.06$ (wave 1), $1.04$ (wave 2), $1.05$ (wave 3), and  $1.09$ (wave 4).

\begin{figure}
\resizebox{12cm}{!}{\includegraphics{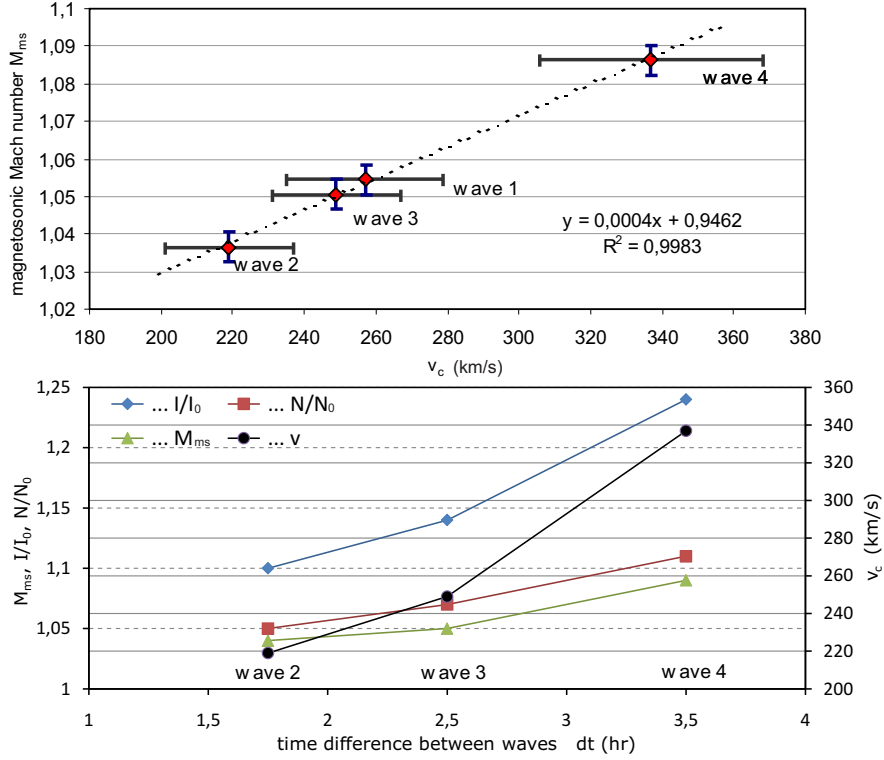}}
\caption{Top: Correlation between magnetosonic Mach numbers $M_{\mathrm{ms}}$, derived from peak perturbation amplitudes, and linear propagation velocities $v_{\mathrm{c}}$ of the four coronal waves. Error bars in the Mach numbers are determined by the uncertainties in the gaussian fits to the perturbation amplitudes (see Fig.~\ref{fig4}, top panels). Bottom: Intensity amplitude ($I/I_0$), density amplitude ($N/N_0$), Mach number ($M_{\mathrm{ms}}$), and wave velocity ($v_{\mathrm{c}}$) presented as a function of the time lags between successive waves.\label{fig5}}
\end{figure}

%Fig. 5
With these observations of homologous waves, we can for the first time perform a quantitative analysis of the characteristic wave parameters without any limiting factors like changing or unknown quiet Sun background conditions. The top panel of Fig.~\ref{fig5} shows the calculated magnetosonic Mach number versus the propagation velocities $v_{\mathrm{c}}$ of the four waves, revealing a distinct correlation between the wave characteristics $M_{\mathrm{ms}}$ and $v_{\mathrm{c}}$ with a correlation coefficient of  $R^2\approx0.99$. Such correlation is expected for nonlinear fast-mode magnetosonic waves because of the relation $M_{\mathrm{ms}} = v_{\mathrm{c}}/v_{\mathrm{ms}}$ \citep[cf.][]{priest82}, where $v_{\mathrm{c}}$ is the coronal wave speed and $v_{\mathrm{ms}} = \sqrt{v_\mathrm{A}^2 + c_\mathrm{s}^2}$ the fast magnetosonic speed [$v_\mathrm{A} \dots$ Alfv\'en velocity, $c_\mathrm{s}$ \dots sound speed].
The mean magnetosonic speed, derived by averaging the four ratios ${v_c/M_{\mathrm{ms}}}$, yields $250$\,km\,s$^{-1}$ and gives with $c_s = 180$\,km\,s$^{-1}$ (T $\sim$1.5~MK) a mean Alfv\'en velocity of $v_\mathrm{A} = 175$\,km\,s$^{-1}$. Applying one-- to five--fold Saito coronal density models for a propagation height of $\sim 0.1 R_{\bigodot}$, we derive for the low corona during solar minimum conditions a magnetic field strength between $1.5-3.5$\,Gauss.

The magnetosonic Mach numbers of all four waves peak at values less than 1.10 and drop off towards the linear regime ($M_{\mathrm{ms}}\approx1$) as the waves expand. These small magnetosonic Mach numbers together with the correlated wave velocities support the view that each observed coronal wave is a low-amplitude MHD fast-mode wave. Since $M_{\mathrm{ms}}$ is even at the maximum close to the linear regime, this also implies that the fast-mode MHD waves are expected to experience only minor deceleration. Since $v_{\mathrm{c}}$ decreases proportional to $M_{\mathrm{ms}}$, this implies a $< 10\%$ change in velocity for the waves under study. For the strongest wave 4 this corresponds to a value of $\sim30$~km~s$^{-1}$. As the error in velocity is of the same order of $\sim10\%$  ($\sim\pm30$~km~s$^{-1}$ for wave 4), such weak deceleration is hidden in the measurement uncertainties. Finally, we note a clear correlation between the lags between successive waves, interpreted as the build-up times of the magnetic energy for the following wave event, and the intensity $I/I_0$, density $N/N_0$, magnetosonic Mach number $M_{\mathrm{ms}}$, and the propagation velocities $v_{\mathrm{c}}$ of waves 2 to 4 (cf. Fig.~\ref{fig5}, bottom).

In our study of the homologous wave events it was possible to compare the wave pulse characteristics and analyze correlations between the wave parameters, knowing that the four large-scale waves propagate in similar coronal conditions. This means that any found correlation is independent from the indeterminated but constant coronal conditions. Our results provide strong support that the observed large-scale coronal waves are indeed fast-mode magnetosonic waves, and additionally suggest a dependence of the wave parameters upon the build-up time of the magnetic energy.

\acknowledgments
I.W.K., A.M.V., and N.M. acknowledge the Austrian Science Fund (FWF): P20867-N16. The European Community's Seventh Framework Programme (FP7/2007-2013) under grant agreement no. 218816 (SOTERIA) is acknowledged (B.V., M.T.). We thank the STEREO/SECCHI teams for their open data policy.

\bibliographystyle{apj}
%\bibliography{waves1_bib}

\end{document}